\documentclass[aps,prl,superscriptaddress,twocolumn,notitlepage]{revtex4-1} 

\usepackage[dvips]{graphicx}
\usepackage{dcolumn}
\usepackage{bm}
\usepackage{amsmath,amsthm,amssymb}
\usepackage[sort&compress]{natbib}
\usepackage{color}
\usepackage{hyperref}
\usepackage{braket}
\hypersetup{colorlinks=true,linkcolor=blue,citecolor=blue,urlcolor=blue}

\begin{document}

\title{Relaxation of an isolated dipolar-interacting Rydberg quantum spin system}

\author{A. Pi\~neiro Orioli}
\email{pineiroorioli@thphys.uni-heidelberg.de}
\affiliation{Institut f\"ur Theoretische Physik, Universit\"at Heidelberg, Philosophenweg 16, 69120 Heidelberg, Germany}

\author{A. Signoles}
\email{signoles@physi.uni-heidelberg.de}
\affiliation{Physikalisches Institut, Universit\"at Heidelberg, Im Neuenheimer Feld 226, 69120 Heidelberg, Germany.}

\author{H. Wildhagen}\thanks{Present address: Department of Neuro- and Sensory Physiology, University Medical Center G\"ottingen, Germany}
\affiliation{Physikalisches Institut, Universit\"at Heidelberg, Im Neuenheimer Feld 226, 69120 Heidelberg, Germany.}

\author{G. G\"unter}
\affiliation{Physikalisches Institut, Universit\"at Heidelberg, Im Neuenheimer Feld 226, 69120 Heidelberg, Germany.}

\author{J. Berges}
\affiliation{Institut f\"ur Theoretische Physik, Universit\"at Heidelberg, Philosophenweg 16, 69120 Heidelberg, Germany}
\affiliation{ExtreMe Matter Institute EMMI, Planckstra\ss e 1, 64291 Darmstadt, Germany}

\author{S. Whitlock}
\affiliation{IPCMS (UMR  7504) and ISIS (UMR  7006),
University of Strasbourg and CNRS, 67000 Strasbourg, France}
\affiliation{Physikalisches Institut, Universit\"at Heidelberg, Im Neuenheimer Feld 226, 69120 Heidelberg, Germany.}

\author{M. Weidem\"uller}
\affiliation{Physikalisches Institut, Universit\"at Heidelberg, Im Neuenheimer Feld 226, 69120 Heidelberg, Germany.}
\affiliation{Hefei National Laboratory for Physical Sciences at the Microscale and Department of Modern Physics, and CAS Center for Excellence and Synergetic Innovation Center in Quantum Information and Quantum Physics, University of Science and Technology of China, Hefei, Anhui 230026, China.}

\begin{abstract}

How do isolated quantum systems approach an equilibrium state? We experimentally and theoretically address this question for a prototypical spin system formed by ultracold atoms prepared in two Rydberg states with different orbital angular momenta. By coupling these states with a resonant microwave driving we realize a dipolar XY spin-1/2 model in an external field. Starting from a spin-polarized state we suddenly switch on the external field and monitor the subsequent many-body dynamics. Our key observation is density dependent relaxation of the total magnetization much faster than typical decoherence rates. To determine the processes governing this relaxation we employ different theoretical approaches which treat quantum effects on initial conditions and dynamical laws separately. This allows us to identify an intrinsically quantum component to the relaxation attributed to primordial quantum fluctuations.
\end{abstract}

\maketitle

\textit{Introduction.} A many-body quantum system initially prepared in an out-of-equilibrium state can produce beautifully complex dynamics as a consequence of strong interparticle interactions. Usually, these same interactions also imply coupling to uncontrolled environmental degrees of freedom, causing rapid relaxation of the system towards an equilibrium state~\cite{Breuer2002}. Remarkably though, relaxation is not unique to open quantum systems. Even when effectively isolated from any external environment, an effective loss of details about initial conditions can occur for relevant observables of the unitary quantum many-body dynamics~\cite{Rigol2008,Polkovnikov2011,Eisert2015}. The mechanisms by which this occurs are difficult to disentangle, but their identification could have important implications for quantum technologies aiming to exploit many-body coherences and may motivate more efficient theoretical approaches for tackling many-body quantum dynamics.

Ultracold atoms, ions and molecules offer unique platforms for addressing such questions due to their high degree of isolation from their environments and the possibility to realize prototypical models such as quantum spin models~\cite{Micheli2006,Bloch2012,Blatt2012}. They are also particularly suited to studying non-equilibrium dynamics such as collective spin dynamics~\cite{Lanyon2011,Simon2011,Martin2013,Barredo2015,Labuhn2016}, spin squeezing~\cite{Gross2010,Riedel2010,Hamley2012,Muessel2014,Bohnet2016}, propagation of correlations~\cite{Cheneau2012,Fukuhara2013,Langen2013,Jurcevic2014,Richerme2014} and relaxation or localization effects~\cite{Trotzky2012,Hild2014,Schreiber2015,Marcuzzi2016,Smith2016,Choi2016}. While the vast majority of these experiments have involved either low-dimensional geometries or contact interactions, spin models can also be realized in three dimensions with dipolar interactions (scaling with interparticle distance as $r^{-3}$) by using magnetic atoms~\cite{Lahaye2009,DePaz2013}, polar molecules~\cite{Yan2013,Hazzard2013} or highly excited Rydberg states of atoms~\cite{Gunter2013}. This provides an interesting border case between long- and short-range physics and can dramatically influence the resulting phase structure~\cite{Yi2007}, stability properties~\cite{Lahaye2007,Baier2016,Li2016} and dynamics~\cite{Eisert2013,Yao2014}.

\begin{figure}[tbp!]
  \hspace*{-0.1cm}
      \includegraphics[width=0.95\columnwidth]{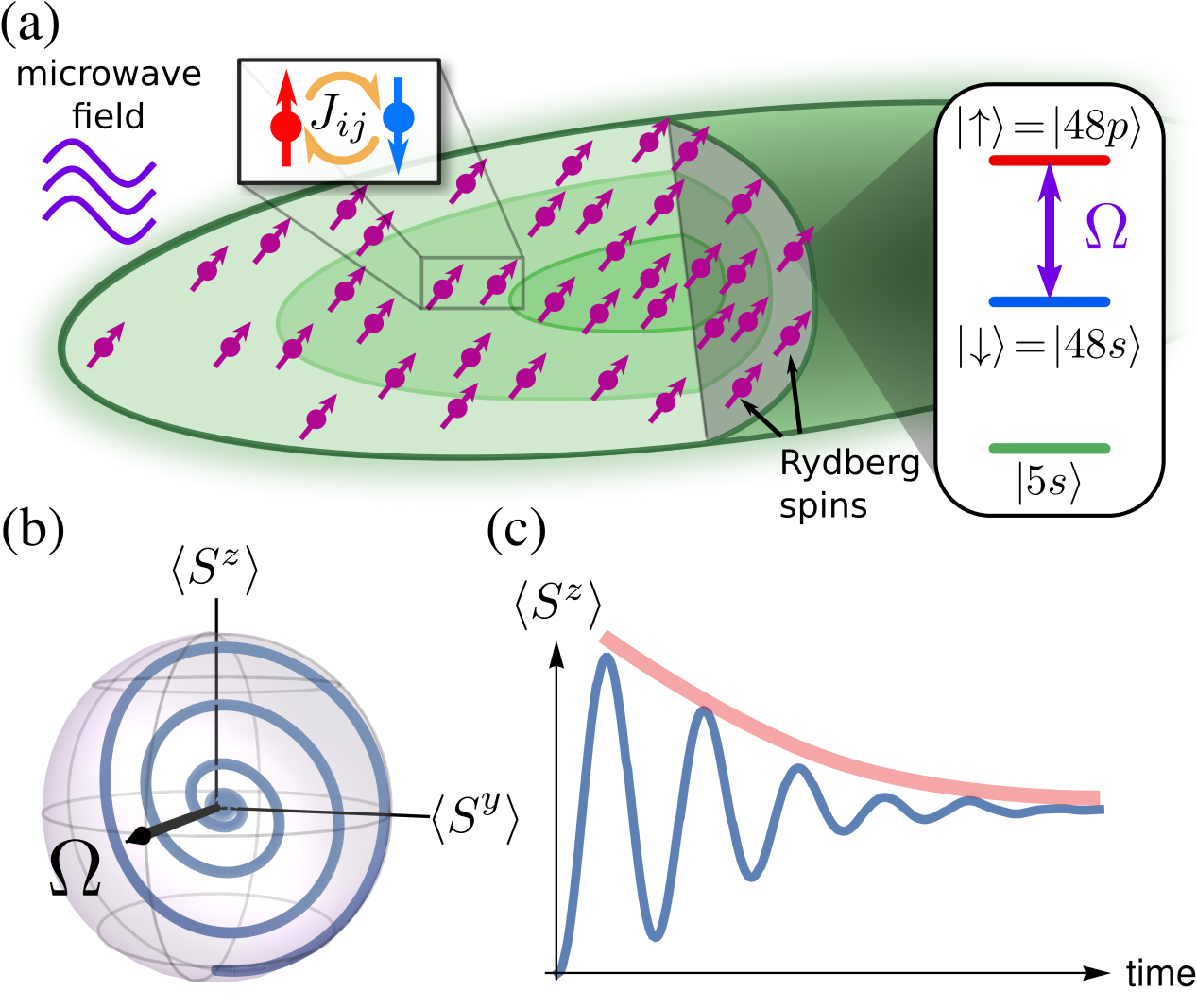}
      \caption{Sketch of a dipolar-interacting Rydberg spin system. (a) In an ultracold atomic gas Rydberg excitations are optically prepared in the initial state $\ket{\downarrow}$. Then a microwave field is turned on which resonantly couples the $\ket{\downarrow}$ and $\ket{\uparrow}$ states. (b) This drives coherent spin oscillations of each atom in the ensemble as seen in the Bloch sphere representation of a single embedded spin, based on the numerical solution of the 8 spin Schr\"odinger equation. (c) Strong dipolar exchange interactions between the spins $J_{ij}$ compete with the driving field leading to relaxation of the spin oscillations.
      }
    \label{fig:exp_sketch}
\end{figure}

In this Letter we experimentally and theoretically investigate the non-equilibrium quantum dynamics of an effectively isolated many-body spin system as it approaches an equilibrium state. Experimentally, we consider a three-dimensional ensemble of ultracold Rydberg atoms, driven by an external microwave field between two states which evolve through coherent and long-range dipolar interactions (Fig.~\ref{fig:exp_sketch}a). Upon suddenly switching on the microwave field we observe oscillatory dynamics and density-dependent relaxation of the many-body system to a practically unmagnetized state. Theoretically, this system can be well described by a dipolar XY quantum spin-1/2 Hamiltonian, which we use to further investigate the origin of the relaxation. Through a well-defined hierarchy of theoretical approximations we disentangle the different mechanisms leading to relaxation of the collective spin in isolated quantum many-body systems, clearly identifying the important role of primordial quantum fluctuations associated with the initial state of the experiment. Finally we show that the many-body dynamics observed in the experiment can indeed be accurately described by unitary quantum dynamics of the dipolar spin-1/2 model without the need to invoke any external decoherence processes.

\begin{figure}[t]
	\hspace*{-0.1cm}
	\includegraphics[width=\columnwidth]{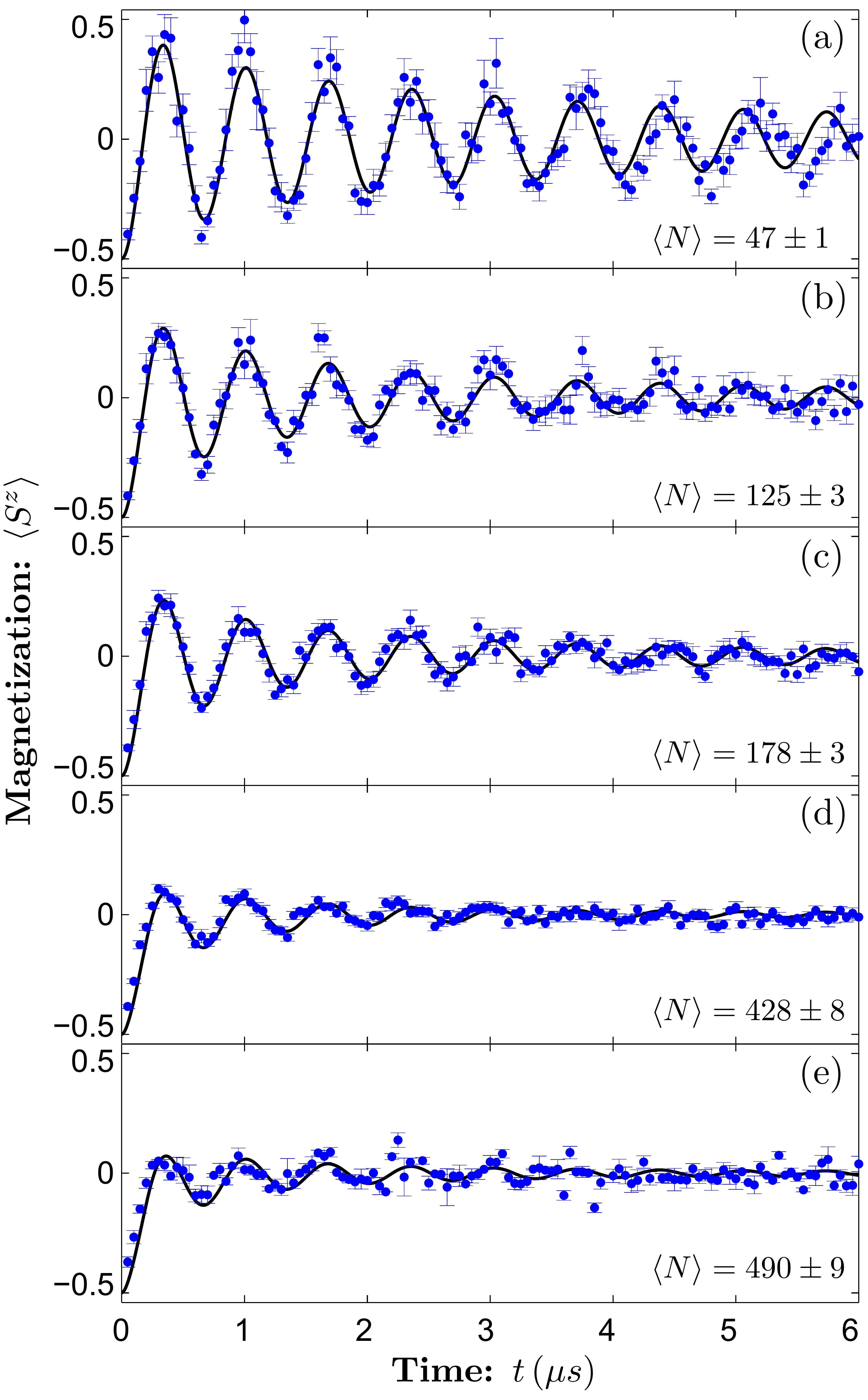}
	\caption{Non-equilibrium magnetization dynamics for various mean numbers of spins $\langle N \rangle$ within a fixed volume. The experimental Rydberg data is represented by blue points. The error bars represent the standard errors of the mean. Also given are results of a MACE simulation for the unitary dynamics of a dipolar XY quantum spin-$1/2$ model starting from a pure state with all spins pointing down (black solid line) as decribed in the text.}
	\label{fig:MACE}
\end{figure}

\textit{Relaxation dynamics in ultracold Rydberg gases.} We perform experiments on a gas of $^{87}\text{Rb}$ atoms prepared in a cigar-shaped optical dipole trap (Fig.~\ref{fig:exp_sketch}a) which constitutes a many-body pseudo-spin-1/2 system. To initialize the system, a small fraction of atoms are prepared in the spin-down state $\ket{\downarrow}=\ket{48S_{1/2},m_j=+1/2}$ by resonant two-photon laser excitation from the electronic ground state. The peak density of spins can be varied between $4\times 10^{7}\,\rm{cm}^{-3}$ and $5\times 10^{8}\,\rm{cm}^{-3}$ by tuning the ground-state density before the excitation pulse. We use a magnetic field of 6.4\,G to isolate a pair of Zeeman states and apply then a microwave field of $35.1\,$GHz to resonantly couple the $\ket{\downarrow}$ and $\ket{\uparrow}=\ket{48P_{3/2},m_j=+3/2}$ states with Rabi frequency $\Omega/2\pi=(1.48\pm 0.05)\,$MHz. The relevant spin-spin interactions originate from the dipole-dipole interactions~\cite{PhysRevA.75.032712,Ravets2015} between pairs of atoms involving different combinations of $\ket{\uparrow},\ket{\downarrow}$.

The experiment starts with $N$ atoms spin-polarized in the state $\ket{\downarrow}$ ($N_\uparrow = 0, N_\downarrow = N$). We then suddenly switch on the microwave field which quenches the system far-from-equilibrium. For vanishing spin density, the dynamics of the magnetization $\langle S^z \rangle=\langle N_\uparrow\rangle / \langle N\rangle - 1/2$ exhibits coherent oscillations between $\pm 1/2$ at the frequency $\Omega$ (Figs.~\ref{fig:exp_sketch}c and \ref{fig:MACE}a). As the density is increased, dipole-dipole interactions compete with the coherent oscillations to produce more complex many-body dynamics. The effect of these interactions can be thought of as causing different atoms in the ensemble to become correlated, ultimately leading to a decaying total magnetization as sketched in Fig.~\ref{fig:exp_sketch}b, c. For all practical purposes the state with $\langle {S}^z \rangle=0$ is the equilibrium state, as for a many-body system the time scale associated with possible revivals grows rapidly with system size. To experimentally measure the magnetization we optically de-pump the $\ket{\downarrow}$ state and then detect the remaining $\ket{\uparrow}$ population by field-ionization which is then averaged over several realizations (further details on the experimental protocol can be found in the Supplementary Material~\cite{SM}). The mean total number of spins $\langle N \rangle$ is inferred by assuming the system equilibrates to a balanced state ($\langle S^z \rangle =0$).

The blue points in Fig.~\ref{fig:MACE} show the time evolution of the magnetization measured for different mean numbers of spins $\langle N \rangle$ within a fixed volume (corresponding to different spin densities). Over the full dataset the magnetization appears to oscillate at a single frequency and with a damping rate between $0.2\,\rm{MHz}$ and $5\,\rm{MHz}$. In particular, the high contrast Rabi oscillations seen in the lowest density data (Fig.~\ref{fig:MACE}a) demonstrate the successful isolation of a two-level system within the Rydberg-state Zeeman manifolds. As the spin density is increased (Fig.~\ref{fig:MACE} from a--e) we observe a transition to an overdamped regime with a damping time which is at least an order of magnitude faster than any single-spin decoherence processes.

\textit{Theoretical description of unitary spin dynamics.} 
Motivated by the experimental results of Fig.~\ref{fig:MACE}, we now seek to understand the mechanisms responsible for the relaxation of isolated quantum spin systems using state-of-the-art many-body theory. As a benchmark system we focus on the dipolar XY spin Hamiltonian, which provides a good description of Rydberg gases involving states which can exchange energy by resonant electric dipole interactions~\cite{SM}. Including the microwave coupling field we arrive at the dipolar XY Hamiltonian in an external field ($\hbar=1$):
 \begin{equation}
 {H}=\frac{1}{4}\sum_{i,j} J_{ij}( S_i^+ S_j^-+ S_i^- S_j^+)+\Omega\sum_i S_i^x \, ,
 \label{eq:H}
 \end{equation}
where $S_i^\alpha$ $(\alpha=\{x,y,z\})$ refers to the spin-1/2 angular momentum operator for spin $i$, $S_i^\pm = (S_i^x \pm \mathrm{i} S_i^y)$ are the spin raising and lowering operators and $J_{ij}$ for $i\neq j$ is the matrix element of the dipole-dipole interaction
\begin{equation}
	{J}_{ij} = \frac{C_3(1-3\cos^2{\theta_{ij}})}{R_{ij}^3} \, .
\label{eq:dipole_int}
\end{equation}
Here $C_3/2\pi=-1.73\,\rm{GHz\,\mu m^{3}}$ characterizes the strength of the exchange interaction proportional to the square of the transition dipole moment connecting the $\ket{\downarrow}$ and $\ket{\uparrow}$ states, $R_{ij}$ is the relative separation between two spins $i$ and $j$ and $\theta_{ij}$ is the angle between their internuclear axis and the quantization axis.

We will assume that the spins are distributed according to a three-dimensional Gaussian distribution with parameters chosen to closely match the conditions of the experiment, although we verify that our results do not depend much on the precise details~\cite{SM}. A spatial configuration of spins is generated by sequentially choosing random coordinates from this distribution and discarding cases which fall within a characteristic distance of $3.8\,\mu$m of another spin to account for the Rydberg blockade effect~\cite{Comparat2010}. From the resulting configuration we then compute the interaction coefficients $J_{ij}$ according to Eq.~\eqref{eq:dipole_int}. For the simulations we average the magnetization over different position realizations until the results converge.

For sufficiently small numbers of spins, the dynamics of (\ref{eq:H}) can be solved without further approximations by direct diagonalization. A corresponding numerical solution for a system of $8$ spins is sketched in Fig.~\ref{fig:exp_sketch}b, c. For larger systems one has to resort to additional approximations, such as those implemented in the so-called Moving-Average Cluster Expansion (MACE) method~\cite{Hazzard2014}. MACE consists of considering appropriately chosen clusters of spins and solving the time-dependent quantum problem within each cluster by diagonalization of the Hamiltonian. The size of the clusters is increased until the results become insensitive to their size. This method is particularly suitable for the calculation of average magnetizations, and it has been successfully used to model the non-equilibrium dynamics of interacting polar molecules in optical lattices~\cite{Hazzard2014,Yan2013}.

Results of the MACE simulations for the unitary evolution of the spin-$1/2$ system at different densities are shown in Fig.~\ref{fig:MACE} (solid lines). One observes a characteristic decay of the magnetization oscillations which is similar to that found in the experiment. However, before comparing theory and experiment in more detail, we will look to identify the main mechanisms responsible for the relaxation of an isolated quantum spin system.

\textit{Role of primordial quantum fluctuations.} 
In general, the theoretical description of a non-equilibrium quantum many-body problem is based on two major ingredients: 
1)~the specification of an \textit{initial state} at time $t=t_0$ and 2)~the \textit{dynamical laws} that evolve this state to later times $t > t_0$. The initial state considered here can be specified either by the full wavefunction $\ket{\downarrow \dots \downarrow}$ or, equivalently, in terms of a set of expectation values of spin operators such as the means $\langle S^\alpha_i(t_0) \rangle$ and variances $\Delta S^{\alpha}_{i}(t_0)=\langle S^\alpha_i(t_0)^2 \rangle - \langle S^\alpha_i(t_0) \rangle^2$ for $\alpha = \{x,y,z\}$. At time $t=t_0$ the non-zero terms characterizing the pure state are
\begin{eqnarray}
	\langle S^z_i(t_0)\rangle =-1/2,~\Delta S^{x}_{i}(t_0)=\Delta S^{y}_{i}(t_0)=1/4.
\label{eq:initial}
\end{eqnarray}
The full quantum Heisenberg equations of motion for the spin system with Hamiltonian (\ref{eq:H}) are given by
\begin{align}
\begin{aligned}
\left( \begin{matrix} \dot{S}_i^x \\ \dot{S}_i^y \\ \dot{S}_i^z \end{matrix} \right) =&\,
\left( \begin{matrix} \Omega + K_i^x \\ K_i^y \\ 0  \end{matrix} \right)
\times
\left( \begin{matrix} S_i^x \\ S_i^y \\ S_i^z  \end{matrix} \right)
\label{eq:eom}
\end{aligned}
\end{align}
where we define $K_i^\alpha \equiv \sum_j J_{ij} S_j^\alpha$.

The question of the origin of relaxation can be approached very efficiently from theory by treating separately the roles of quantum fluctuations in the initial state and in the dynamical laws that describe the non-equilibrium problem. We first consider a mean-field (MF) approximation where quantum fluctuations both in the initial conditions and in the time evolution equations are fully neglected. More precisely, in the MF approach quantum spins are replaced by classical spins initialized to the quantum expectation values $\langle S^\alpha_i(t_0)\rangle$ with no fluctuations, i.e.~$\Delta S^\alpha_i(t_0)= 0$. These classical spins are then evolved with the Heisenberg equations (\ref{eq:eom}) with each spin operator replaced by the corresponding expectation value. With this approach the spin-spin interaction given in Eq.~\eqref{eq:H} is approximated by the interaction of each spin with the averaged field created by all the other spins and the fluctuations of this mean-field are neglected. Therefore, the Hamiltonian (\ref{eq:H}) leads to a closed set of classical evolution equations for the expectation values $\langle S^\alpha_i (t)\rangle$ that are numerically solved using a Crank-Nicholson algorithm. The mean-field computation is performed for the same spin configurations and couplings as for the MACE simulations.

\begin{figure}[tbp!]
  \hspace*{-0.1cm}
      \includegraphics[width=\columnwidth]{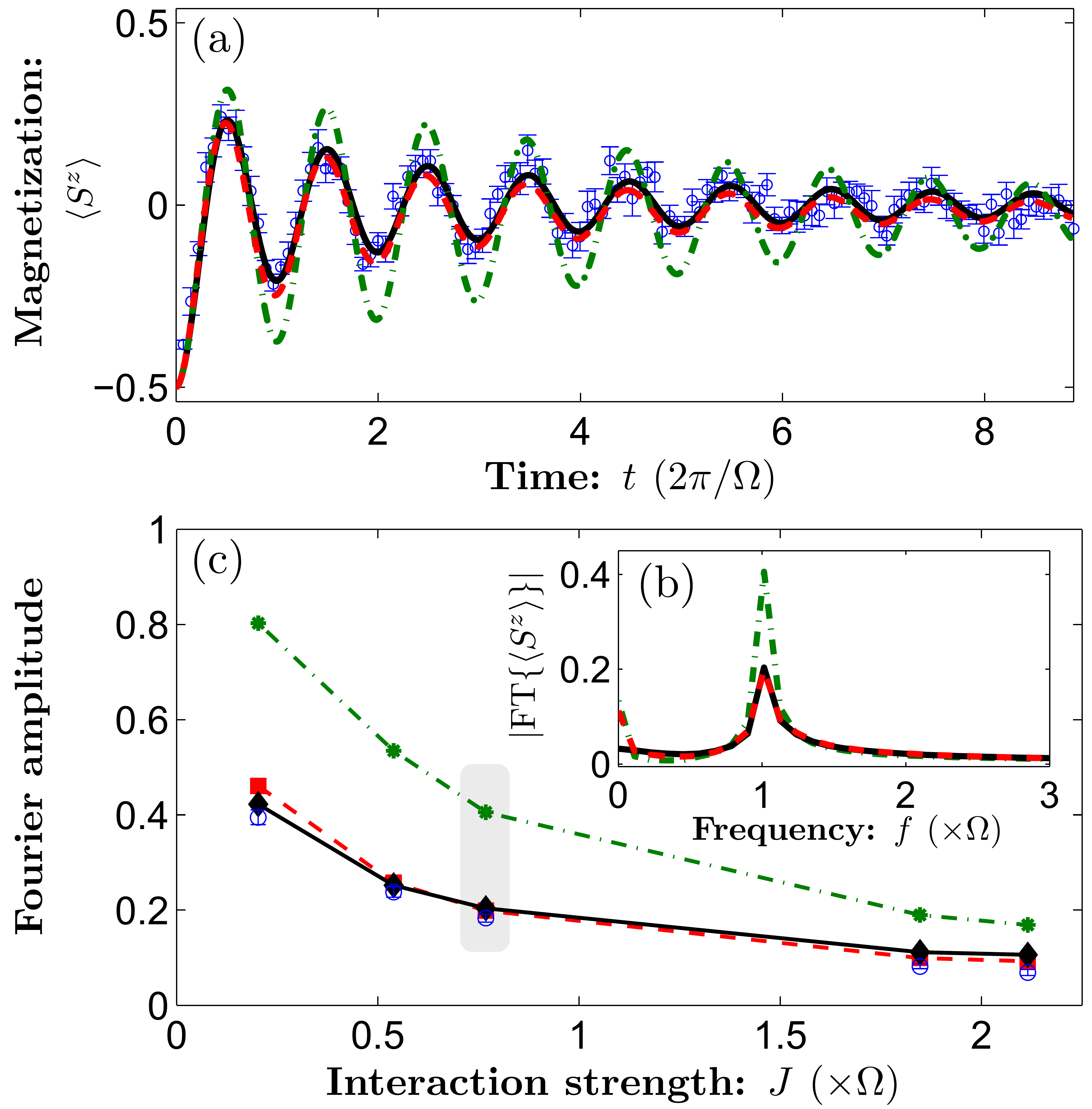}
      \caption{Relaxation dynamics computed by three theoretical methods: MACE (black solid lines), MF (green dash-dotted lines) and TWA (red dashed lines). Experimental data (blue dots) shown for comparison. (a) Magnetization as a function of time for $\langle N \rangle =178$. (b) Amplitude of the Fourier transform of (a) normalized to the value expected for a full contrast oscillation. (c) Peak amplitude of the Fourier component $f=\Omega$ for different interaction strengths $J$. The grey box corresponds to the above parameter choice. Error bars on the experimental data correspond to the standard deviation of the sampling distributions estimated using the bootstrap method.}
    \label{fig:theory}
\end{figure}

Fig.~\ref{fig:theory}a shows the MF prediction of the magnetization (dash-dotted green line) compared to the MACE simulation (solid black line) for an intermediate density corresponding to $\langle N \rangle=178$. In the MF approximation relaxation arises as a consequence of dephasing: due to their random position in the atomic cloud, the spins experience different mean fields and precess at various frequencies, resulting in density-dependent relaxation of the collective spin observables. However, Fig.~\ref{fig:theory}a shows that this dephasing distinctly underestimates the MACE simulation results, demonstrating that MF does not capture all the essential underlying processes and that quantum fluctuations cannot be fully neglected.

In a next step we include quantum fluctuations in the initial state while retaining the classical equations of motion to evolve the system to later times. Therefore, such an approximation takes into account only part of the quantum effects originating from the quantum initial state. This approach is also known as the (discrete) Truncated Wigner Approximation (TWA)~\cite{Polkovnikov2010,PhysRevLett.114.045701,PhysRevA.79.042703,PhysRevX.5.011022}, where the quantum initial state is modelled by an ensemble of classical spins with $\pm1/2$ in the directions transversal to the mean $\langle S_i^\alpha(t_0)\rangle$ that reproduces the quantum fluctuations at $t=t_0$~\cite{PhysRevX.5.011022}. Each member of the ensemble is then evolved using the same classical equations of motion as in the MF approximation and the results are obtained from ensemble averages.

Remarkably, the dynamics of the total magnetization computed with TWA (dashed red line in Fig.~\ref{fig:theory}a for $\langle N \rangle=178$) and MACE (solid black line) are in rather close agreement. To quantify this agreement and to extend it to different interaction strengths, we consider the Fourier transform of the time-evolution of the magnetization (Fig.~\ref{fig:theory}b). We see that apart from a small difference at frequency component $f=0$, the TWA and MACE simulations closely agree, in particular for $f\approx \Omega$. The narrow peak in the Fourier amplitude at $f=\Omega$ corresponds to the single-spin precession frequency and it provides a convenient measure for comparing the different theoretical results (Fig.~\ref{fig:theory}c). The close agreement between TWA and MACE persists for different mean interaction strengths $J=C_3(N/V)^3$, corresponding to the whole range of densities considered in Fig.~\ref{fig:MACE}. In contrast, the MF simulations systematically exhibit about half the damping of the other methods (see the Supplementary Material for further comparisons~\cite{SM}). 

This comparison sheds light onto the role of quantum fluctuations for the relaxation process. While neglecting all fluctuations (MF approximation) greatly underestimates the relaxation rate, the inclusion of just the initial quantum fluctuations of the spins via TWA manages to capture the essential quantitative features of the relaxation dynamics. Since TWA evolves the quantum initial conditions using purely classical dynamics, the similar success obtained with both TWA and MACE indicates that the dynamical fluctuations do not play an important role for the considered collective spin observable (magnetization). In turn, this shows that the dynamics of collective spin observables can probe the effects of primordial quantum fluctuations intrinsic to the spin-1/2 system.

\textit{Comparison to experiment.} We now investigate whether the essential features of the experimental data can indeed be described by unitary dynamics of a spin-$1/2$ system. To make this comparison it is necessary to adjust the effective volume of the spin distribution used in the simulations to the experiment. For this we take the lowest-density experimental data and perform simulations assuming different Gaussian volumes $V$, while keeping the total number of spins fixed, until optimal agreement is found. The obtained volume $V=(2.7\pm 0.4)\times 10^5\,\mu$m$^3$ is kept fixed in all further simulations and different densities are simulated by changing the number of spins in accordance with the experiment. We note that the obtained effective volume is approximately 0.3 times the naive estimate based on the excitation laser beam waist and cloud size. This may be to partially compensate uncertainties in experimental parameters such as the ion detection efficiency.

In Fig.~\ref{fig:MACE} we compare the MACE simulations with the experimental data. Remarkably, the simulated dynamics are in good agreement with the data for the full range of experimental parameters, including the density-dependent damping. A slight discrepancy of the oscillation frequency at long times is seen (e.g.~in Fig.~\ref{fig:MACE}a), which may be explained by transient power fluctuations of the microwave field after switching on the source. As it is not possible to perform the experiment in the absence of dipolar interactions altogether, we cannot completely rule out that the relaxation dynamics seen in Fig.~\ref{fig:MACE}(a) are due to possible sources of additional fluctuations in the experiment. However, the good agreement between the data and the MACE simulations points out that the observed relaxation can indeed be described in terms of unitary evolution of the Hamiltonian (1) and intrinsic quantum fluctuations alone, without the need to invoke any external decoherence processes (for details on the exclusion of other possible noise sources see the Supplementary Material~\cite{SM}). The success of the dipolar XY model to capture the essential features of the experiment shows that the Rydberg spin systems can serve as valuable testbeds for addressing fundamental questions about non-equilibrium quantum many-body dynamics, such as the role of quantum fluctuations.

Comparing the MF predictions to the experimental data in the same way does not allow us to describe essential quantitative aspects such as the full density dependence of the damping, see Figs.~\ref{fig:theory}a and c. This seems to rule out MF as a valid description, whereas TWA and MACE both describe the data equally well, which indicates that primordial quantum fluctuations play a crucial role in the relaxation dynamics. To resolve additional quantum corrections arising through the dynamical laws (which would be at least partially captured by MACE) would require a much higher level of experimental precision or measurements of different observables. For example, higher-order moments of the collective spin distribution may be used to distinguish MACE and TWA simulations.

\textit{Conclusions and outlook.}
We have experimentally and theoretically investigated the many-body relaxation of the magnetisation in a prototypical quantum spin system following a quantum quench. Dipolar interacting Rydberg spin systems are ideally suited to studying the nature of correlations that emerge as a consequence of many-body quantum dynamics. Our analysis reveals the role of disorder and the importance of the initial quantum fluctuations on the relaxation process. While mean field models often perform well for describing the general properties of systems with long-range interactions~\cite{Mori2010}, here we have shown for three-dimensions and $1/r^3$ interactions that primordial quantum corrections play a crucial role for the non-equilibrium dynamics. The study can be generalized to other observables and initial conditions, which could also be accessed in experiment (e.g., by full counting statistics~\cite{schempp2014,malossi2014}).

\acknowledgments{We acknowledge valuable discussions with Ignacio Aliaga Sirvent, Titus Franz, Martin G\"arttner, Thomas Gasenzer, Vladislav Gavryusev, Stephan Helmrich, Ana Mar\'ia Rey, Arghavan Safavi-Naini and Johannes Schachenmayer. This work is part of and supported by the DFG Collaborative Research Centre "SFB 1225 (ISOQUANT)" and by the Heidelberg Center for Quantum Dynamics, the European Union H2020 FET Proactive project RySQ (grant N. 640378) and the Deutsche Forschungsgemeinschaft under WH141/1-1.}\\

A.P.O. and A.S. contributed equally to this work.

\bibliography{spindynamics_bibliography}

\end{document}